AI-Assisted Curation of Conference Scholarship: Compiling, Structuring, and Analyzing Two Decades of Presentations at the Society for Social Work and Research (2005–2026)

Brian E. Perron, Bryan G. Victor, and Zia Qi

**Brian E. Perron**
School of Social Work, University of Michigan

**Bryan G. Victor**
School of Social Work, Wayne State University

**Zia Qi**
School of Social Work, University of Michigan

**Author Note**

Correspondence concerning this article should be addressed to Brian E. Perron, School of Social Work, University of Michigan, 1080 S. University Ave., Ann Arbor, MI 48109. Email: bperron@umich.edu

**Abstract**

**Purpose:** This study developed a comprehensive database of presentation abstracts from the Society for Social Work and Research (SSWR) Annual Conference and examined patterns in research methodology, authorship, collaboration, and institutional participation over two decades.

**Method:** Abstract metadata was compiled from the SSWR Confex conference management system for presentations from 2005 to 2026 using web scraping. A small language model (gpt-oss:20b) performed classification and extraction tasks on abstracts, including categorization of methodologies and parsing of author affiliations, with human review at each major stage to ensure accuracy.

**Results:** The database contains 23,793 presentations with 69,924 author records representing 20,779 unique researchers from 4,049 institutions across 93 countries. Annual conference presentations increased from 423 in 2005 to 1,935 in 2026, representing a compound annual growth rate of 7.5%. Quantitative methods predominated (61.1%), followed by qualitative approaches (23.4%), mixed methods (9.1%), and reviews (5.4%). The mean number of authors per presentation increased from 2.22 in 2005 to 3.31 in 2026. International participation grew from 4.5% to 13.5% of author affiliations over the observation period.

**Discussion:** Findings indicate substantial growth in SSWR conference participation, alongside increased collaboration and international engagement. The methodological distribution reveals continued quantitative predominance with growing qualitative representation. This database provides research infrastructure for systematic hypothesis testing about research priorities and disciplinary development over time, enabling analyses that inform both scholarship and conference planning.

**AI-Assisted Curation of Conference Scholarship: Compiling, Structuring, and Analyzing Two Decades of Presentations at the Society for Social Work and Research (2005–2026)**

Scientific conferences occupy a distinctive position in the research lifecycle. They are rarely the endpoint of scholarly work but often its first public articulation, in which emerging ideas, methods, and findings are shared before formal peer-reviewed publication. As such, conference presentations provide an important indicator of where a discipline is investing its intellectual labor and research attention, who is participating, and where scholars might be innovating methodologically. Patterns in conference scholarship can anticipate future publication trends, reveal lines of inquiry that may never progress to journal publication, and illuminate the training and participation of early-career researchers whose work may be underrepresented in the formal publication record.

These considerations are particularly salient for social work. The Society for Social Work and Research (SSWR) Annual Conference, the primary research-focused conference for social work in North America, has hosted tens of thousands of presentations since its inception in 1994 (see Williams et al., 2008; Fong et al., 2019). Collectively, these presentations represent an extraordinary volume of scholarly activity, encompassing study design, data collection, analysis, and interpretation across the full range of social work research domains. Yet unlike some scientific fields in which conference proceedings or abstracts are formally published and indexed, social work conference presentations are disseminated exclusively through conference programs designed for event navigation rather than analysis. Abstracts persist as unstructured digital records, with authors, affiliations, methodological approaches, and institutional participation embedded in unstructured text fields rather than standardized metadata in an analyzable database. As a result, despite the central role of the SSWR conference in North American social work scholarship, the field lacks a systematic account of how this scholarship and conference participation have evolved.

For purposes of scientometric and bibliometric inference in social work, conference abstracts offer a particular advantage, capturing a point when scholarly activity is still consolidated within the field before it is dispersed across disciplinary and extra-disciplinary journals. Prior research indicates that social work scholars commonly publish in extra-disciplinary outlets, particularly in areas such as health, aging, violence, and substance use, where relevant literatures span multiple fields (Green & Baskind, 2007; Hodge & Yu, 2025). Recent evidence also suggests that a majority of tenure-track faculty in PhD-conferring programs prioritize submitting their highest-quality work to journals outside their discipline, often because of the research content rather than disciplinary affiliation (Hodge & Yu, 2025). As a result, journal-based analyses of social work scholarship, particularly those limited to disciplinary outlets, necessarily underrepresent the full scope of research conducted by social work scholars. While the SSWR Annual Conference is one of several venues for social work scholarship—the Council on Social Work Education's Annual Program Meeting remains the field's largest conference—SSWR's research emphasis brings a substantial concentration of the field's research activity together in a single venue, offering an opportunity to observe patterns in empirical scholarship before it is distributed across multiple disciplinary literatures.

However, despite their analytic potential, SSWR conference abstracts have rarely been used in empirical studies of social work scholarship. We are aware of only three peer-reviewed studies that have systematically analyzed SSWR conference abstracts, each limited to a subset of available presentations (see Kang et al., 2025; Perron et al., 2011; Perron & Qi, 2025). Perron and colleagues (2011) examined publication rates among SSWR presenters, finding that approximately 41% of presentations were published in peer-reviewed journals within several years. Kang and colleagues (2025) analyzed the carbon footprint of conference attendance, raising sustainability questions as meeting size increases. Perron and Qi (2025) used SSWR abstracts to examine engagement with postmodern and critical theoretical frameworks, demonstrating differential adoption across methodological traditions. This stands in contrast to the extensive scientometric and bibliometric literature examining social work scholarship through disciplinary journals (e.g., Victor et al., 2017), where standardized metadata and indexing facilitate large-scale analysis. Victor and colleagues (2017) documented the rise of co-authorship in social work journal publications from 1989 to 2013, finding that collaboration was associated with higher citation impact. We suspect that the relative absence of conference-based analyses reflects not a lack of scholarly relevance, but more likely the practical constraints imposed by unstructured abstract formats and the absence of analyzable metadata.

Fortunately, recent advances in artificial intelligence offer a pathway for curating a population-level, analyzable database of SSWR abstracts and their associated metadata. Large language models have demonstrated remarkable capabilities in natural language understanding, entity extraction, and text classification tasks that previously required extensive human annotation (Brown et al., 2020). However, deploying such models has typically required substantial computational infrastructure, proprietary application programming interface (API) access, or cloud-based processing, raising concerns about cost, sustainability, and environmental impact. The energy consumption and carbon footprint associated with training and deploying large-scale models are growing concerns as models expand in size (Strubell et al., 2019; Patterson et al., 2021).

A countervailing trend toward smaller, more efficient language models challenges assumptions that performance scales primarily with model size. Research has demonstrated that models with far fewer parameters can achieve competitive performance on domain-specific tasks while dramatically reducing computational requirements (Hinton et al., 2015; Sanh et al., 2019). These small language models (SLMs), typically defined by parameter counts in the single-digit to low double-digit billions, as opposed to frontier models with hundreds of billions to trillions of parameters, offer particular advantages for applied research contexts. Because of their reduced computational requirements, SLMs are small enough to run on consumer-grade hardware without cloud dependencies, which in turn offers benefits such as data privacy and lower computational costs. They can be deployed locally to maintain data privacy. For tasks where SLM performance is sufficient, local deployment can substantially reduce computational costs relative to cloud-based frontier models and can be adapted to specialized extraction tasks through targeted prompting. For disciplines operating under resource constraints, SLMs can democratize access to computational text analysis that was previously available only to well-funded computer science laboratories or through expensive commercial services.

The present study leverages these technologies to systematically compile, structure, and analyze two decades of SSWR conference presentations. We used a locally hosted small language model to extract structured metadata from unstructured abstract text and author affiliation strings, thereby transforming an informational archive into usable scholarly infrastructure. We build on preliminary work that validates an AI-driven classification methodology using SSWR conference data. Perron and Qi (2025) demonstrated reliable multi-model LLM classification for analyzing theoretical and methodological content, achieving substantial inter-rater agreement across six state-of-the-art language models. The present investigation extends that foundation with three objectives. First, we document methods for constructing a comprehensive database of SSWR presentations from 2005 through 2026, including procedures for web scraping, content extraction, and language model-based classification. Second, we examine longitudinal trends in conference activity, analyzing patterns in presentation volume, research methodology, and geographic participation. Third, we chart trends in authorship, collaboration, and career-stage representation.

<div align="center">**Methods**</div>

**Data Source and Scope**

The SSWR Annual Conference has utilized the Confex conference management platform since the mid-2000s, with presentation abstracts, author information, and session metadata publicly accessible through structured web pages. This study compiled metadata from all available conference years, spanning the 9th Annual Conference (2005) through the 30th

Annual Conference (2026). Abstracts before 2005 were not publicly available at the time of this study. For reference, we present our full processing pipeline as Figure 1.

[INSERT FIGURE 1 ABOUT HERE]

Conference programs featured multiple formats: oral presentations, poster sessions, symposia, and roundtable discussions. We retained all presentation formats that contained accompanying abstracts, as format category boundaries varied across conference years and all represented substantive research contributions subject to competitive peer review. Individual presentations within symposia and roundtable sessions were included, though high-level overview descriptions for these sessions were excluded, retaining only the component presentations. Workshops and keynote lectures were excluded because they either did not report original scholarship or did not undergo the standard peer-review process. Presentations lacking abstract text or with abstracts shorter than 50 characters (n = 67) were also excluded, as these contained only titles without substantive content.

**Data Collection**

Data collection used a multi-stage pipeline that combined web scraping and language model processing. Web scraping is the automated process of retrieving information from publicly accessible web pages for collection and analysis in a structured format. The primary data source was the SSWR Confex archive, accessed via systematic HTTP requests. We developed custom scraping scripts that first retrieved conference program index pages listing all sessions, then systematically accessed individual presentation pages to extract titles, abstracts, and author information. We compared extracted record counts against conference program indices for each year as an ongoing quality check and believe the database captures the full universe of publicly available abstracts in the Confex system. The webscraping was performed using Crawl4AI (UncleCode, 2024). All data processing was implemented in Python (version 3.11).

HTML parsing was performed with BeautifulSoup4 (version 4.12; Richardson, 2004) using the lxml parser backend to extract structured content from web pages. HTML parsing involves reading the underlying code of a web page so that specific pieces of information, such as titles, abstracts, and author details can be systematically identified and extracted. For each presentation, we extracted the title from header elements, the abstract from designated content divisions, and author information from affiliation blocks. Session metadata, including the presentation format (oral, poster, symposium, or roundtable) and the session title, were captured when available.

The HTML structure of conference pages evolved across the 22-year observation period, requiring year-specific parsing logic to accommodate formatting changes. Conference years 2005–2008 used a different URL structure and page layout than subsequent years, requiring specialized extraction procedures. In the later years (2009–2026), the HTML structure was more consistent but still showed variations in underlying formatting. We implemented iterative validation by comparing extracted records against manually reviewed samples to ensure parsing accuracy across all conference years.

**Variable Extraction and Normalization**

To extract structured metadata from unstructured affiliation text, we used a multi-stage pipeline that combines language-model parsing, normalization algorithms, and entity resolution. Author information presented the most substantial extraction challenge due to inconsistent formatting across presentations and years. The Confex system, used to serve each conference website, stores author affiliations as semi-structured text strings combining names, credentials, positions, and institutional information in variable formats. For example, a single affiliation field might contain "Matthew Smith, Professor, School of Social Work, University of Michigan, Ann Arbor, MI" or simply "M. Smith, U. Michigan" depending on what authors entered during submission.

To extract structured information from these heterogeneous strings, we used the gpt-oss-20b language model (OpenAI, 2025) to parse author affiliations. This is a small language model and can be run locally without reliance on proprietary cloud-based application programming interfaces. The selection of gpt-oss-20b was motivated by performance, computational efficiency, and environmental sustainability, given its lower energy consumption compared with larger proprietary models (Qi et al., 2025). The model was deployed using llama.cpp, software that enables local deployment of language models.

We designed structured prompts directing the model to extract standardized fields from raw affiliation text, outputting results in JSON format with the following schema: author name, academic degrees (e.g., PhD, MSW, LCSW), position/title, institution name, department, city, state/province, and Country/Region. Representative prompts and detailed extraction schemas are provided in the Supplementary Material. The prompt included few-shot examples demonstrating the expected output format for various affiliation string patterns common in academic social work contexts. The model was configured with a temperature of 0.1 to minimize output variability and ensure consistent extraction across similar input patterns. Context length was set to 8,192 tokens, which is sufficient to process affiliation strings alongside instructional prompts. To validate parsing accuracy, we manually reviewed a random sample of 200 parsed records and found that the model correctly extracted institutions in 94% of cases and positions in 91% of cases.

**Author Names.** Author names were normalized using the NFKD (Normalization Form Compatibility Decomposition) algorithm from Python's unicodedata module. This process decomposes characters into their base forms, removes combining marks, and converts accented characters to ASCII equivalents (e.g., "é" to "e," "ñ" to "n"). Whitespace was normalized by collapsing multiple spaces into a single space and removing leading and trailing spaces. Asterisks marking presenting authors were removed. The resulting normalized names were stored in a separate field for matching while preserving the original casing in the primary name field.

To identify unique researchers across presentations, we implemented a graph-based entity resolution system combining multiple similarity metrics. Name parsing used the HumanName class from the nameparser library to extract structured components (first name, last name, middle initial, suffix). We used a blocking strategy that grouped names by first initial and last name to reduce computational complexity, comparing only names within the same block rather than all possible pairs.

Within each block, name similarity was calculated using a multi-factor scoring approach. Base similarity used the token *sort-rati*o function from the RapidFuzz library (Bachmann, 2020), which computes fuzzy string similarity after sorting tokens alphabetically (handling variations such as "Smith, John" vs. "John Smith"). Special handling was applied to common Asian surnames (e.g., Lee, Kim, Park, Chen, Wang, Liu, Zhang) because these surnames are highly prevalent; exact first-name matches were required rather than fuzzy matching, preventing false-positive matches among unrelated researchers sharing common surnames.

Additional factors modified the base similarity score, including matching middle initials (for Western names only), shared institutional affiliations, and logical progression of positions over time. Name pairs achieving similarity scores of 90 or higher were connected as edges in a NetworkX graph. This threshold was empirically derived through sensitivity testing, balancing the trade-off between matching accuracy and comprehensive record linkage. Connected components in this graph identified clusters of name variants belonging to the same individual. For each cluster, a canonical name was selected using a heuristic preferring longer names with more name parts (e.g., "Michael S. Sherraden" preferred over "Mike Sherraden" or "M. Sherraden"). The entity resolution process reduced 23,481 raw unique name variants to 20,779 canonical names across 69,924 author-presentation records.

**Institutions.** The language model extracted the primary academic or organizational affiliation from each author record; authors listing multiple affiliations were assigned to the first-listed institution. Because the model extracted institutions as they appeared in affiliation text, variants such as "University of Michigan," "U. Michigan," "U-M," "UM School of Social Work," and "University of Michigan-Ann Arbor" required consolidation. We used a language model to help create a mapping table of canonical institution names and their associated variants, covering the most frequently occurring institutions. The normalization pipeline first cleaned institution strings by removing embedded position titles and geographic suffixes (state abbreviations, Country/Region names, city-state patterns). Matching proceeded through three stages: exact match against canonical forms, substring matching for variants with department prefixes or suffixes, and pattern matching for common naming variations. Institutions that did not match any canonical form were retained in their cleaned but unstandardized form. We did not make any changes to the author institutions in the original data. Thus, the data set reflects the author's institution at the time of the conference, and not their current institution if they have subsequently switched universities.

**Countries**. Country/Region assignment followed explicit mention in the affiliation text or inference from state/city information (e.g., "Ann Arbor, MI" mapped to the USA). Country/Region names were then normalized to standardized English forms using a mapping dictionary addressing over 80 Country/Region variations. Examples include: "United States of America," "US," "U.S.," "U.S.A.," and "America" all mapped to "USA"; "Republic of Korea" and "Korea" mapped to "South Korea"; "P.R. China," "People's Republic of China," and "Mainland China" mapped to "China"; "UK," "Britain," and "England" mapped to "United Kingdom." State and province abbreviations within the United States and Canada were expanded to full names (e.g., "MI" to "Michigan," "ON" to "Ontario") to ensure consistency across records.

**Positions.** The language model extracted position categories, which were then normalized using a standardized taxonomy through keyword-based classification with regular expression pattern matching. The LLM-extracted position field was searched for indicator terms in priority order: "doctoral student," "PhD student," "PhD candidate," or "ABD" triggered classification as Doctoral Student; "assistant professor" as Assistant Professor; "associate professor" as Associate Professor; "professor" (without assistant/associate modifiers) as Full Professor; "clinical professor," "professor of practice," or "teaching professor" as Clinical Professor; "postdoc," "post-doctoral," or "postdoctoral" as Postdoctoral; leadership roles were differentiated between academic settings (Senior Leadership Academic: deans, associate deans, department chairs) and practice settings (Senior Leadership Practice: executive directors, program directors, agency administrators); "research associate," "research scientist," "project coordinator," or "research assistant" as Research Staff; "research professor" or "research associate professor" as Research Faculty; "lecturer" or "instructor" as Instructor; "adjunct" as Adjunct Faculty; "MSW student," "master's student," or "graduate student" (without doctoral indicators) as Masters Student; "BSW student," "undergraduate," or "bachelor's" as Undergraduate Student. Positions not matching any category were classified as Practitioner or marked as Unknown when position information was missing or unparseable.

**Methodology Classification**

Research methodology was classified for each presentation based on an analysis of the abstract content using the same gpt-oss:20b model. We implemented a classification scheme distinguishing five primary categories: quantitative (studies using statistical analysis of numeric data), qualitative (studies employing interpretive analysis of textual, visual, or observational data), mixed methods (studies explicitly integrating quantitative and qualitative approaches), review (systematic reviews, scoping reviews, and meta-analyses synthesizing existing literature), and theoretical/other (conceptual papers, methodological discussions, and presentations not fitting other categories).

The classification prompt instructed the model to analyze each abstract for methodological indicators, including data type descriptions (surveys, interviews, administrative records), sample characteristics, analytical methods mentioned (regression, thematic analysis, content analysis), and explicit methodology statements. Detailed classification prompts and category definitions are provided in the Supplementary Material. The prompt specified that classification should be based on the dominant methodology when multiple approaches were mentioned, unless the abstract explicitly described integrating quantitative and qualitative components (triggering mixed methods classification).

To validate model performance, we manually classified a random stratified sample of 60 abstracts (10 per methodology category) and calculated inter-rater agreement between human and model classifications. Cohen's kappa was .83, indicating almost perfect agreement according to standard benchmarks (Landis & Koch, 1977). Classification accuracy varied by category: quantitative (100%), qualitative (100%), review (90%), mixed methods (80%), and theoretical/other (75%). The lower accuracy for theoretical/other reflects the heterogeneous nature of this residual category, which includes simulation studies, policy descriptions, methodological tutorials, and implementation case studies.

**Human Review Procedures**

The study team completed iterative quality review at each major stage of the data processing pipeline. Human review occurred after webscraping to verify the completeness of extracted records, after language model extraction to assess the accuracy of parsed affiliation fields, after entity resolution to examine author deduplication clusters, and after methodology classification to validate classification assignments. Review criteria included the accuracy of extracted fields, the consistency of normalized values, and the logical coherence of entity clusters. The pipeline was human-supervised throughout and should not be interpreted as fully automated.

**Data Analysis**

All statistical analyses were conducted using Python with the pandas library (version 2.0) for data manipulation and aggregation. Descriptive statistics characterized the database's composition, including temporal coverage, counts of presentations by year, distribution of methodologies, and metadata completeness. We calculated compound annual growth rate (CAGR) using the standard formula, where $n$ represents the number of years. Growth analyses examined the total number of presentations and the mean number of authors per presentation over the observation period.

$$CAGR = [(\frac{Ending\ Value}{Beginning\ Value})^{(1/n)}] - 1$$

Collaboration patterns were assessed using authorship metrics, including mean and median number of authors per presentation, the proportion of multi-authored versus single-authored presentations, and trends in team size over time. Geographic distribution was assessed at the Country/Region level based on author affiliations, with international participation defined as the proportion of author records affiliated with non-US institutions.

**Ethical Considerations**

This study analyzed publicly available conference abstracts and author information from the SSWR Confex archive, accessible without authentication, login credentials, or other access restrictions. No confidential or protected information was collected. All materials analyzed had been voluntarily submitted by authors for public dissemination as part of the conference program. The study involved no interaction with human subjects and was therefore exempt from institutional review board oversight. Data collection procedures followed established norms for responsible web scraping, including rate limiting and delays between HTTP requests to minimize server load and avoid disruption of access for other users.

<div align="center"><strong>Results</strong></div>

**Dataset Overview**

The final dataset contained 23,793 presentations and 69,924 author records from the SSWR Annual Conferences, spanning 2005 through 2026 (Table 1). Author records represent individual author-presentation associations; the same researcher presenting multiple times, or multiple authors on a single presentation, each generates a separate record. Normalized name matching identified 20,779 unique authors. Authors were affiliated with 4,049 unique institutions across 93 countries.

[INSERT TABLE 1 ABOUT HERE]

**Conference Growth**

Conference presentations increased substantially over the observation period, from 423 presentations in 2005 to 1,935 in 2026 (See Figure 2). This growth represents a compound annual growth rate of 7.5%, indicating that SSWR presentations roughly doubled every decade. The growth trajectory exhibited distinct phases: relatively stable output from 2005–2011, averaging 488 presentations annually; rapid expansion from 2012–2016, averaging 1,466 presentations; and sustained high volume from 2017–2026, averaging 1,566 presentations. The 2015 conference marked a notable increase in volume, with presentations increasing from 801 in 2014 to 1,166 in 2015, a 46% increase in a single year. The 2026 conference had the highest single-year total (n = 1,935) in conference history.

[INSERT FIGURE 2 ABOUT HERE]

**Methodology Trends**

Quantitative methods predominated across the observation period, accounting for 61.1% of all presentations (n = 14,542), followed by qualitative approaches (23.4%, n = 5,565), mixed methods (9.1%, n = 2,172), reviews (5.4%, n = 1,274), and theoretical/other (1.0%, n = 240). This distribution demonstrates continued quantitative emphasis in social work research while indicating substantial representation of interpretive methodological traditions.

Temporal analysis revealed gradual methodological diversification (see Figure 3). Quantitative presentations declined from approximately 71% of the program in early years to 51% by 2025–2026, while qualitative presentations increased from 15% to 30% over the same period. Mixed-methods research showed modest fluctuations over time, with presentations remaining roughly 9-11%. Review methodologies have increased from negligible representation in the early years (i.e., 1.7% in 2005) to approximately 8% of recent programs, reflecting a growing emphasis on evidence synthesis. The theoretical/other category remained relatively stable at 1–2% throughout the observation period.

[INSERT FIGURE 3 ABOUT HERE]

**Authorship and Collaboration**

The mean number of authors per presentation increased from 2.22 in 2005 to 3.31 in 2026 (see Figure 4), indicating a shift toward larger research teams over time. Single-authored presentations declined from 38% of the 2005 program to 21% in 2026, while presentations with four or more authors increased from 16% to 38%.

[INSERT FIGURE 4 ABOUT HERE]

**Position Distribution**

Analysis of author roles (Table 2) indicates that Assistant Professors (19.6%, n = 13,726) and Doctoral Students (19.0%, n = 13,312) were the most frequent contributors, collectively accounting for nearly 39% of all author records. Associate Professors (13.1%, n = 9,182), Research Staff (12.5%, n = 8,713), and Full Professors (12.3%, n = 8,569) also maintained substantial representation. Position information was unavailable or unparseable for 8.1% of records (n = 5,686). Senior Leadership roles, split between practice settings (3.0%) and academic settings (0.8%), collectively account for 3.8% of author records. The Practitioner category (1.2%) captures social workers, consultants, physicians, and non-academic professionals, reflecting the scope of conference participation by the broader social work practice community and allied health professionals.

Academic positions among first authors showed shifting trends within faculty ranks, with doctoral students maintaining a robust and consistent presence as primary presenters throughout the observation period (see Figure 5). Among faculty, Assistant Professors have historically represented the largest share of presenters; however, in recent years, their share has declined. Conversely, Associate Professors have shown a gradual upward trend, while Full Professors have remained relatively stable, with a slight increase in recent years.

[INSERT FIGURE 5 ABOUT HERE]

**Geographic Distribution**

Authors from institutions in the United States dominated conference participation, accounting for 89.1% of author affiliations (n = 61,691; Table 3). Canada represented the second-largest national presence at 3.7% (n = 2,586), followed by South Korea (2.1%, n = 1,451), Israel (0.9%, n = 613), and Hong Kong (0.7%, n = 490). The remaining 88 countries collectively contributed 4.5% of author affiliations. International participation, defined as the proportion of author affiliations from non-US institutions, increased from 4.5% in 2005 to 13.7% in 2026 (see Figure 6). This trend reflects growing global engagement with the SSWR conference, with notable acceleration after 2018.

[INSERT FIGURE 6 ABOUT HERE]

<div align="center">

**Discussion**

</div>

This study established a comprehensive database of SSWR Annual Conference presentations spanning two decades, enabling systematic analysis of research trends, authorship patterns, and institutional participation at the discipline's premier research venue. Taken together, these findings invite reflection at two levels: what patterns in conference participation reveal about SSWR as a scholarly space, and what they suggest about broader trends in social work scholarship.

We view the creation of a comprehensive, longitudinal database of SSWR metadata as a primary contribution of this study. The database functions as research infrastructure that supports the generation and testing of hypotheses about how research priorities, methodological traditions, and theoretical engagement develop over time within social work scholarship. For example, Perron and Qi (2025) used a subset of SSWR conference abstracts to examine temporal patterns in social work researchers' engagement with postmodern and critical theoretical frameworks, demonstrating that increases in theoretical alignment were not evenly distributed across methodologies but instead concentrated within qualitative research. Analyses of this kind illustrate how conference-based datasets can provide empirical leverage on long-standing debates about theory, method, and knowledge production. To maximize scholarly utility, we are pursuing a phased approach to data release, with a focus on accessibility and validation. Following beta testing with social work researchers to support quality review and error identification, we anticipate a public release of the database with accompanying documentation, hosted on a sustainable platform with version control to support longitudinal updates as future conferences occur.

At the conference level, participation in SSWR has expanded dramatically over the past two decades, with particularly rapid growth in recent years. The volume of presentations increased from 423 annually in 2005 to 1,935 in 2026, representing a compound annual growth rate of 7.5%. The 2015 conference marked a particularly notable inflection point, with presentations increasing 46% in a single year (from 801 to 1,166). This growth may reflect multiple contextual factors, including expansion of doctoral programs in social work, growth in SSWR membership, changes in conference programming or submission structures, and broader increases in research funding and institutional support for scholarship. However, given the nature of our data, we cannot establish causal relationships between these factors and conference growth, and multiple mechanisms likely operate simultaneously.

An expansion of conference volume certainly carries benefits. A larger program can accommodate a broader range of topics, methodological approaches, and substantive perspectives, increasing opportunities for participation across institutional types and areas of substantive interest. SSWR functions as a scholarly space for disseminating pre-publication research, building collaborative networks, and supporting early-career development. Sustaining these functions at the current scale requires ongoing attention to factors such as managing program density, supporting mentoring infrastructure, and evaluating presentation formats to maximize both accessibility and scholarly exchange

At the same time, scale introduces potential constraints. Research on conference participation suggests that as programs grow denser, attention can become more fragmented and access to high-value exchanges more unevenly distributed, often favoring those with greater familiarity or social capital within the field (Hauss, 2021). Conferences can also include implicit expectations about navigation and participation that are not equally accessible to all attendees (Caro-Diaz et al., 2024). These general concerns about conference scale align with the substantial presence of doctoral students and assistant professors documented in this study. SSWR has implemented structured mentoring programs alongside conference growth, including student-focused and early-career networking events and dedicated sessions to facilitate connections and professional development. We view these initiatives as well-justified responses to the conference's scale and demographic composition, helping ensure that expansion does not dilute the developmental and community-building functions central to the meeting's purpose.

Participation analyses in this study also indicate that SSWR has increasingly functioned as a site of global scholarly exchange, reflecting the conference's expanding transnational networks and intellectual reach. However, the 2026 conference data reveal a notable decline in international first-author participation relative to prior years. This shift may signal diminished scholarly interest, but we believe it more likely reflects the fragility of global engagement. Heightened geopolitical tensions, visa delays, rising travel costs, and uncertainty about travel to the United States may serve as structural barriers that disproportionately affect international scholars' ability to participate fully. If SSWR is to continue serving as a genuinely cross-national venue for social work scholarship, attention to how global participation is enabled or constrained by forces beyond the conference itself will remain an important consideration.

At the field level, our findings offer insight into broader trends in social work scholarship, particularly among U.S.-based researchers. The increased representation of qualitative research suggests growing recognition of the value of interpretive, contextualized, and meaning-centered approaches for understanding complex social phenomena (Padgett, 2017). This is particularly important in social work, given that qualitative inquiry is well-suited to examining constructs such as lived experience, power, and process, central knowledge for effective and ethical social work practice. This pattern mirrors trends documented in journal-based scientometric analyses. For example, while specific methodological distributions vary by outlet and time period, the broader movement toward methodological pluralism appears consistent across conference and journal scholarship.

Additionally, our findings suggest that the growth of qualitative work has coincided with increased collaboration. This pattern is consistent with trends observed in the journal literature. Victor et al. (2017) documented the rise of co-authorship in social work publications from 1989 to 2013, finding that collaborative work became increasingly prevalent and was associated with higher citation impact. Our findings suggest this collaborative trend extends to conference scholarship as well. Prior research suggests that collaborative or team science can enhance methodological rigor and scholarly impact (Wuchty et al., 2007). However, collaboration is not without accompanying challenges related to coordination, authorship norms, and the evaluation of individual scholarly contributions (Leahey, 2016). Empirical analyses of scholarly collaboration also suggest that expanding collaborative networks entails trade-offs, particularly for early-career scholars, for whom increases in the number of collaborators may fragment time and attention and reduce the sustainability of ongoing collaborative relationships (Wu et al., 2024).

**Limitations**

Several limitations constrain the interpretation of these findings. First, although the entity resolution system employed multiple similarity metrics, author disambiguation remains imperfect. The blocking strategy may miss matches where names are recorded with different initials. Second, the methodology classification relied on abstract content, which may not fully reflect the actual research methods employed. Third, the database captures accepted presentations, not submitted proposals, preventing analysis of acceptance rates. Fourth, presentation formats varied across years, complicating longitudinal comparison, particularly during the virtual/hybrid conferences of 2020/2021. Fifth, the database is limited to abstracts publicly available through the Confex archive at the time of data collection. We do not have access to conference data prior to 2005. Finally, while the database reflects published program content and includes some presentations that were subsequently withdrawn but remain in the Confex system, we cannot fully account for all withdrawals or assess their impact on participation patterns during periods of external disruption, such as the COVID-19 pandemic or geopolitical tensions affecting the 2026 conference.

## Conclusion

Scientific conferences offer a unique window into the early formation of scholarly work, capturing research activity before it is sorted across journals and disciplines. In social work, the SSWR Annual Conference consolidates a substantial share of the field's research output, yet its programs have rarely been treated as empirical data. This study demonstrates how small language models can convert unstructured conference records into analyzable metadata, enabling longitudinal analysis of research activity, collaboration, and methodological change. The resulting database provides a durable infrastructure for examining how social work research is organized, produced, and shared over time. By making conference scholarship visible at scale, this work supports more deliberate reflection on the field's research practices and priorities.

**Table 1 Dataset Summary Statistics**

| Metric | Value |
|---|---|
| Total Abstracts | 23,793 |
| Total Author Records* | 69,924 |
| Unique Authors | 20,779 |
| Unique Institutions | **4,049** |
| Unique Countries/Regions | 93 |
| Years Covered | 22 (2005-2026) |

*Note.* *Author records reflect all author-presentation associations. Unique authors determined via normalized name matching.

**Table 2 Distribution of Author Roles by Academic Position**

| Academic Position | Total (N) | % of Total | First Author % | Co-Author % |
|---|---|---|---|---|
| Assistant Professor | 13,726 | 19.6% | 45.0% | 55.0% |
| Doctoral Student | 13,312 | 19.0% | 45.3% | 54.7% |
| Associate Professor | 9,182 | 13.1% | 32.3% | 67.7% |
| Research Staff | 8,713 | 12.5% | 20.7% | 79.3% |
| Full Professor | 8,569 | 12.3% | 19.0% | 81.0% |
| Unknown | 5,686 | 8.1% | 44.0% | 56.0% |
| Masters Student | 2,674 | 3.8% | 25.3% | 74.7% |
| Senior Leadership Practice | 2,127 | 3.0% | 12.8% | 87.2% |
| Postdoctoral | 1,473 | 2.1% | 45.3% | 54.7% |
| Research Faculty | 1,151 | 1.6% | 23.8% | 76.2% |
| Practitioner | 837 | 1.2% | 16.1% | 83.9% |
| Instructor | 631 | 0.9% | 32.6% | 67.4% |
| Clinical Professor | 616 | 0.9% | 29.4% | 70.6% |
| Senior Leadership Academic | 545 | 0.8% | 15.0% | 85.0% |
| Adjunct Faculty | 489 | 0.7% | 37.2% | 62.8% |
| Undergraduate Student | 193 | 0.3% | 7.3% | 92.7% |

*Note.* Total N = 69,924. Percentages in First Author and Co-Author columns represent the proportion of each position serving in that role (row-level percentages). Column totals: First Author N = 23,791; Co-Author N = 46,133. Position categories normalized using language model classification of raw text. Senior Leadership Academic includes deans, associate deans, and department chairs; Senior Leadership Practice includes executive directors, program

directors, and agency administrators. Research Staff includes research scientists, project coordinators, and research associates. Research Faculty includes research professors and research associate professors. Clinical Professor includes professors of practice and teaching professors. Practitioner includes social workers, consultants, physicians, and non-academic professionals. Unknown indicates records where position information was missing or unparseable.

Table 3. *Top 20 Countries/Regions by Author Role*

| Rank | Country/Region | Total Records | First Author (N) | Co-Author (N) |
|------|----------------|---------------|------------------|---------------|
| 1 | USA | 61,691 | 21,420 | 40,271 |
| 2 | Canada | 2,586 | 791 | 1,795 |
| 3 | South Korea | 1,451 | 521 | 930 |
| 4 | Israel | 613 | 217 | 396 |
| 5 | Hong Kong | 490 | 186 | 304 |
| 6 | China | 452 | 145 | 307 |
| 7 | United Kingdom | 242 | 48 | 194 |
| 8 | Uganda | 236 | 7 | 229 |
| 9 | Taiwan | 203 | 87 | 116 |
| 10 | Kazakhstan | 114 | 4 | 110 |
| 11 | Australia | 98 | 28 | 70 |
| 12 | Switzerland | 80 | 26 | 54 |
| 13 | South Africa | 59 | 8 | 51 |
| 14 | Finland | 56 | 13 | 43 |
| 15 | India | 52 | 6 | 46 |
| 16 | Ghana | 44 | 5 | 39 |
| 17 | Chile | 43 | 9 | 34 |
| 18 | Colombia | 42 | 6 | 36 |
| 19 | Mexico | 39 | 5 | 34 |
| 20 | Singapore | 38 | 16 | 22 |

*Note: Ranked by total author records.* Overall N = 69,924. Entries reflect institutional affiliation locations and include both sovereign states and administrative regions.

**Figure 1**

*Data Processing Pipeline of Presentations at the Society for Social Work and Research Annual Meeting (2005-2026).*

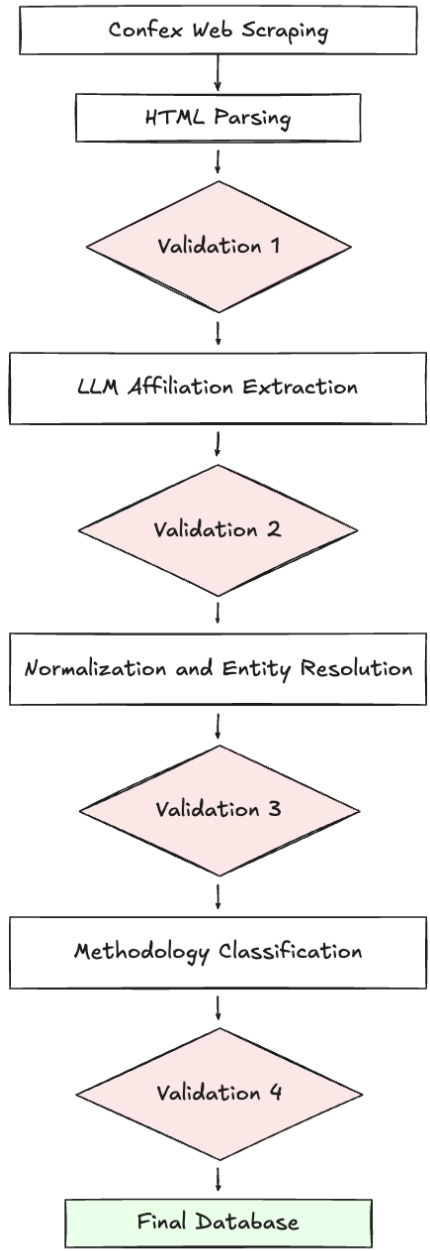

**Figure 2**

*Growth Trends in Scholarship at the Society for Social Work and Research Annual Meeting (2005–2026).*

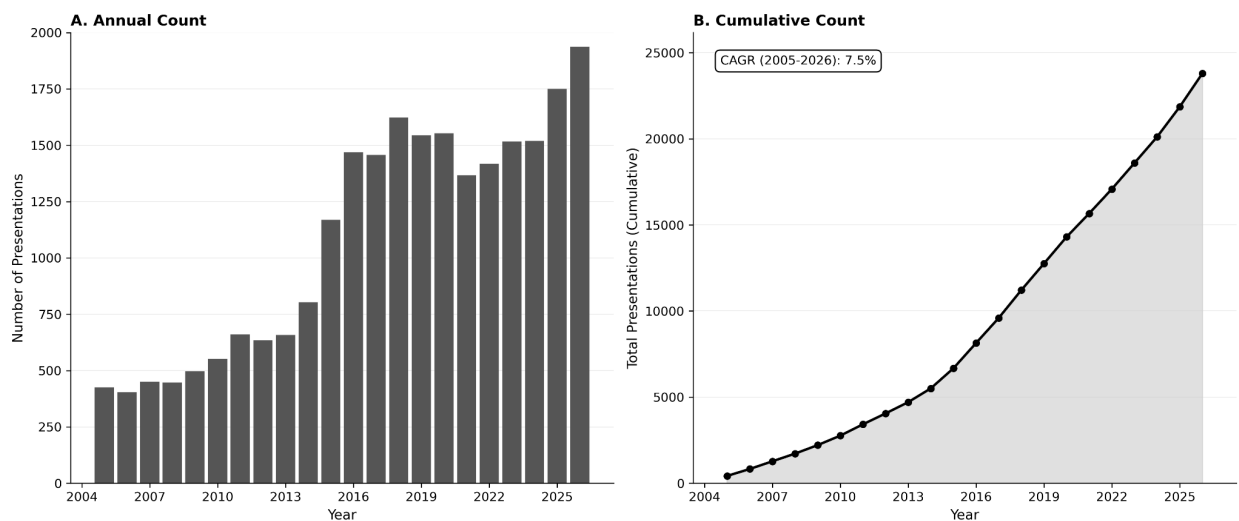

*Note.* The compound annual growth rate (CAGR) indicates a sustained expansion of the field's knowledge base. N = 23,793 presentations.

**Figure 3**

*Research Methodology Trends at the Society for Social Work and Research Annual Meeting (2005–2026).*

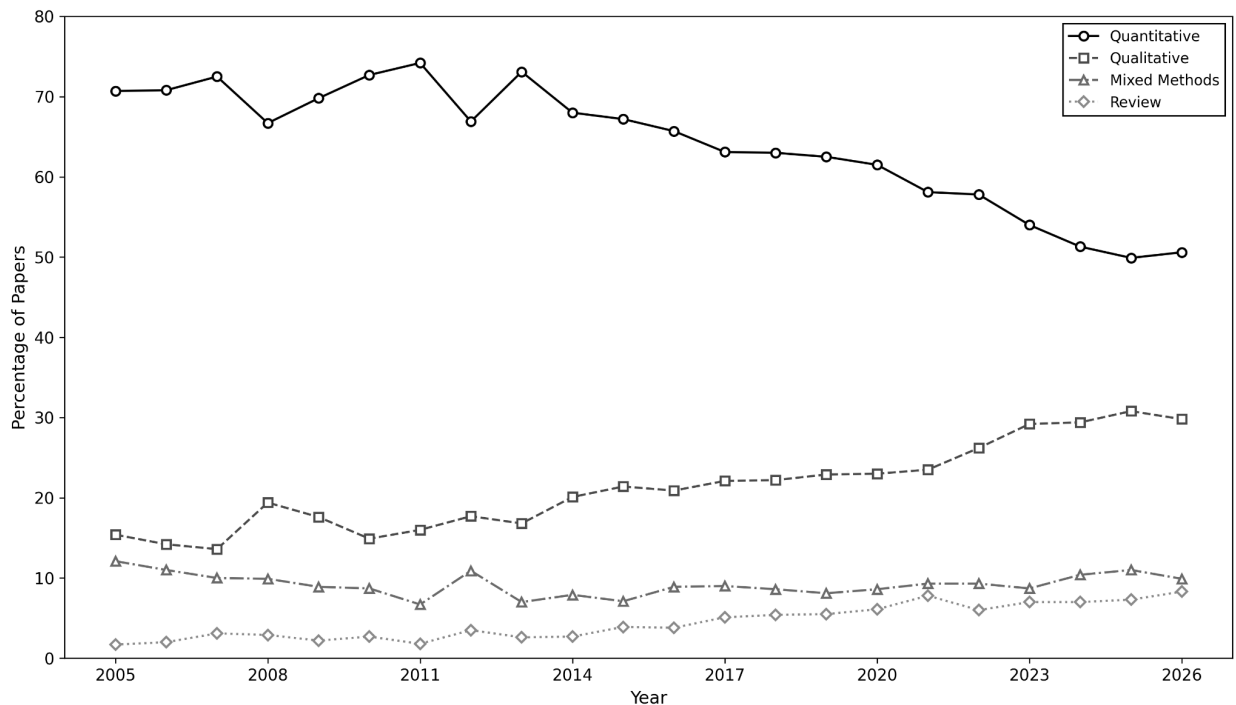

*Note.* N = 23,793 presentations classified using a language model (gpt-oss:20b).

**Figure 4**

*Co-Authorship Trends at the Society for Social Work and Research Annual Meeting (2005–2026).*

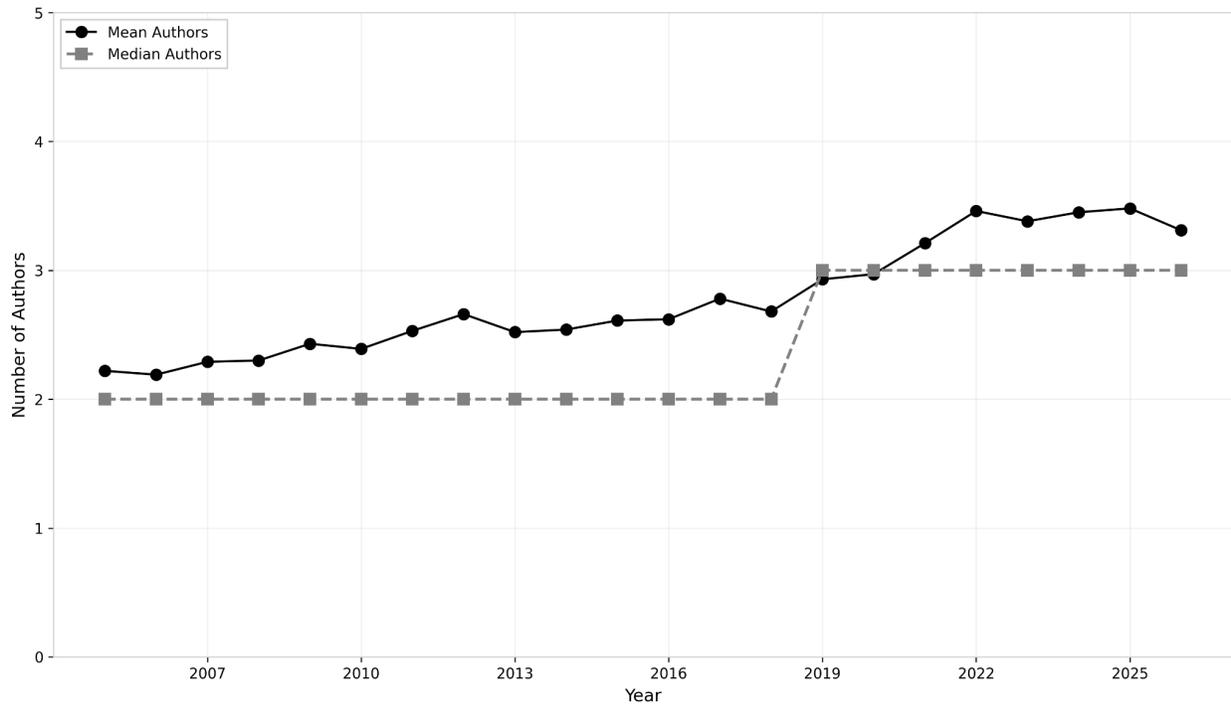

Note: Trends assessed across 23,793 total presentations.

**Figure 5.**

*Academic Career Stage Distribution of First Authors at the Society for Social Work and Research Annual Meeting (2009–2026).*

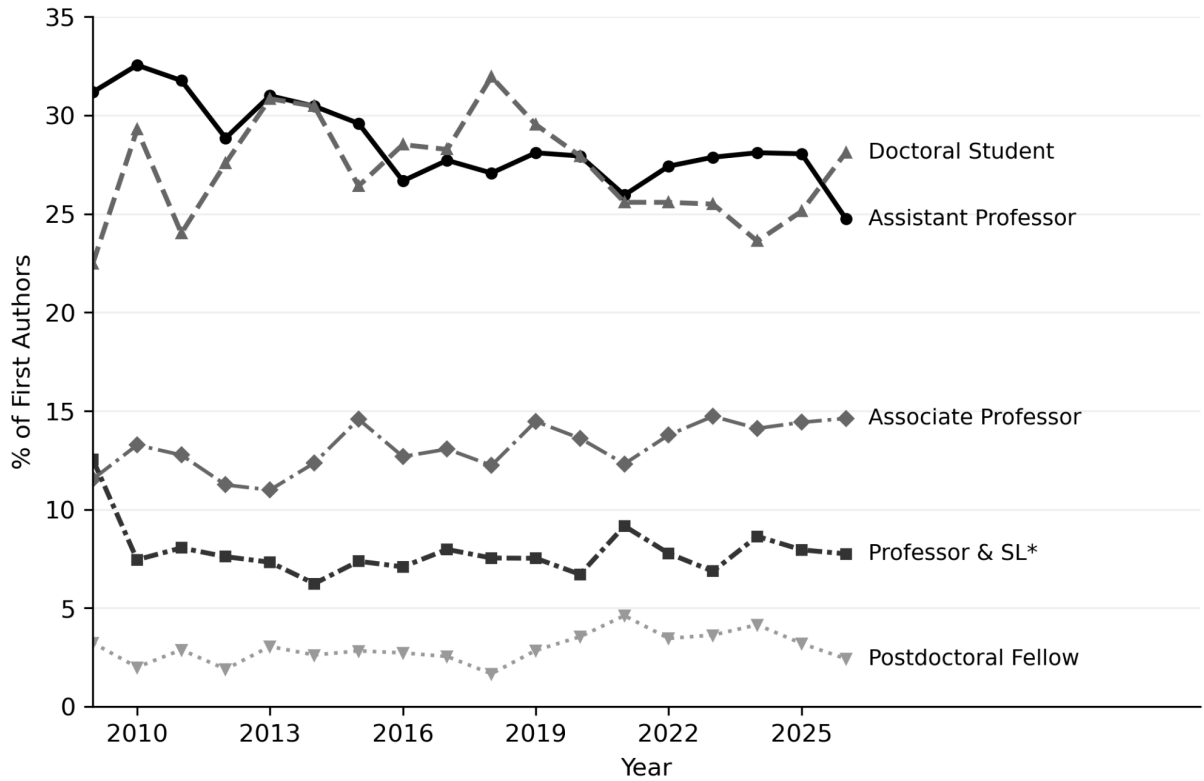

Note: Data for 2005-2008 excluded due to missing position information (>99% unknown). Analysis includes 22,075 first authors from 2009-2026. SL= Senior Leadership (academic).

**Figure 6**

*International Participation Trends at the Society for Social Work and Research Annual Meeting (2005–2026).*

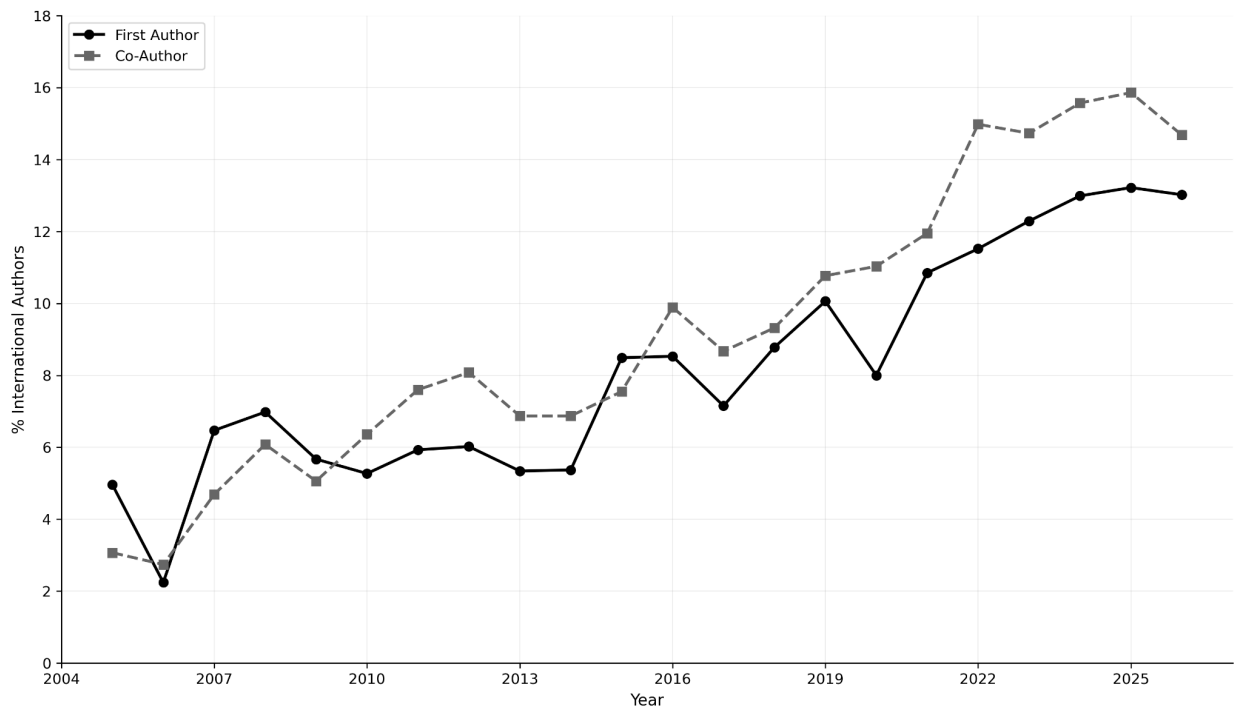

Note: N = 69,924 author records.